\documentclass[12pt]{article}
\usepackage{epsfig}
\usepackage{graphicx}
\usepackage{amsmath}

\begin{document}

 \newcommand{\beq}{\begin{equation}}
\newcommand{\eeq}{\end{equation}}
\newcommand{\bea}{\begin{eqnarray}} 
\newcommand{\eea}{\end{eqnarray}}
\newcommand{\beqn}{\begin{eqnarray}}
\newcommand{\eeqn}{\end{eqnarray}}
\newcommand{\beas}{\begin{eqnarray*}}
\newcommand{\eeas}{\end{eqnarray*}}
\newcommand{\defi}{\stackrel{\rm def}{=}}
\newcommand{\non}{\nonumber}
\newcommand{\bquo}{\begin{quote}}
\newcommand{\enqu}{\end{quote}}
\newcommand{\qt}{\tilde q}
\newcommand{\m}{\tilde m}
\newcommand{\trho}{\tilde{\rho}}
\newcommand{\tn}{\tilde{n}}
\newcommand{\tN}{\tilde N}
\newcommand{\gsim}{\lower.7ex\hbox{$\;\stackrel{\textstyle>}{\sim}\;$}}
\newcommand{\lsim}{\lower.7ex\hbox{$\;\stackrel{\textstyle<}{\sim}\;$}}


\def\de{\partial}
\def\Tr{ \hbox{\rm Tr}}
\def\const{\hbox {\rm const.}}  
\def\o{\over}
\def\im{\hbox{\rm Im}}
\def\re{\hbox{\rm Re}}
\def\bra{\langle}\def\ket{\rangle}
\def\Arg{\hbox {\rm Arg}}
\def\Re{\hbox {\rm Re}}
\def\Im{\hbox {\rm Im}}
\def\diag{\hbox{\rm diag}}


\def\QATOPD#1#2#3#4{{#3 \atopwithdelims#1#2 #4}}
\def\stackunder#1#2{\mathrel{\mathop{#2}\limits_{#1}}}
\def\stackreb#1#2{\mathrel{\mathop{#2}\limits_{#1}}}
\def\Tr{{\rm Tr}}
\def\res{{\rm res}}
\def\Bf#1{\mbox{\boldmath $#1$}}
\def\balpha{{\Bf\alpha}}
\def\bbeta{{\Bf\beta}}
\def\bgamma{{\Bf\gamma}}
\def\bnu{{\Bf\nu}}
\def\bmu{{\Bf\mu}}
\def\bphi{{\Bf\phi}}
\def\bPhi{{\Bf\Phi}}
\def\bomega{{\Bf\omega}}
\def\blambda{{\Bf\lambda}}
\def\brho{{\Bf\rho}}
\def\bsigma{{\bfit\sigma}}
\def\bxi{{\Bf\xi}}
\def\bbeta{{\Bf\eta}}
\def\d{\partial}
\def\der#1#2{\frac{\d{#1}}{\d{#2}}}
\def\Im{{\rm Im}}
\def\Re{{\rm Re}}
\def\rank{{\rm rank}}
\def\diag{{\rm diag}}
\def\2{{1\over 2}}
\def\ntwo{${\mathcal N}=2\;$}
\def\nfour{${\mathcal N}=4\;$}
\def\none{${\mathcal N}=1\;$}
\def\ntwot{${\mathcal N}=(2,2)\;$}
\def\ntwoo{${\mathcal N}=(0,2)\;$}
\def\x{\stackrel{\otimes}{,}}

\def\ba{\beq\new\begin{array}{c}}
\def\ea{\end{array}\eeq}
\def\be{\ba}
\def\ee{\ea}
\def\stackreb#1#2{\mathrel{\mathop{#2}\limits_{#1}}}

\def\Tr{{\rm Tr}}
\newcommand{\cpn}{CP$(N-1)\;$}
\newcommand{\wcpn}{wCP$_{N,\tilde{N}}(N_f-1)\;$}
\newcommand{\wcpd}{wCP$_{\tilde{N},N}(N_f-1)\;$}
\newcommand{\vp}{\varphi}
\newcommand{\pt}{\partial}
\newcommand{\ve}{\varepsilon}
\renewcommand{\theequation}{\thesection.\arabic{equation}}

\setcounter{footnote}0

\vfill

\begin{titlepage}

\begin{flushright}
FTPI-MINN-13/44, UMN-TH-3318/13\\
Jan 6/2014
\end{flushright}

\vspace{1mm}

\begin{center}
{  \Large \bf  
\boldmath{Quantum Deformation of the Effective \\ Theory on Non-Abelian string\\[2mm]
 
and 2D-4D correspondence}}

\vspace{5mm}

 {\large \bf    M.~Shifman$^{\,a}$ and \bf A.~Yung$^{\,\,a,b}$}
\end {center}

\begin{center}

$^a${\it  William I. Fine Theoretical Physics Institute,
University of Minnesota,
Minneapolis, MN 55455, USA}\\
$^{b}${\it Petersburg Nuclear Physics Institute, Gatchina, St. Petersburg
188300, Russia
}
\end{center}

\vspace{1mm}

\begin{center}
{\large\bf Abstract}
\end{center}

We explore non-Abelian strings in the $r=N-1$ vacuum of  \ntwo supersymmetric QCD
with the gauge group U$(N)$ and $N_f$ flavors of quarks ($N_f\geq N$), where $r$ is the number of 
condensed quarks.  \ntwo supersymmetry
is broken down to \none by a small mass
term for the adjoint matter. We discover that the low-energy two-dimensional theory on the string world-sheet
receives nonperturbative corrections from the bulk, through the bulk gaugino condensate. This is in 
contradistinction
with the $r=N$ vacuum situation, in which nonperturbative effects on the world sheet are determined by
internal dynamics of the world-sheet theory. 
The 2D-4D correspondence (the coincidence of spectra of two-dimensional kinks and four-dimensional monopoles)
remains valid in the BPS sector. 
Nonperturbative bulk effects deforming  the weighted CP model on the world sheet 
are found by virtue of the method of resolvents suggested by Gaiotto, Gukov 
and Seiberg for surface defects \cite{GGS}. In the $r=N$ 
vacuum the gaugino condensate in the bulk vanishes, and there are
no ``outside" nonperturbative corrections on the world sheet.

\vspace{2cm}

\end{titlepage}

 \newpage



\section {Introduction }
\label{intro}
\setcounter{equation}{0}

Non-Abelian strings \cite{HT1,ABEKY,SYmon,HT2} were first found in \ntwo supersymmetric QCD,
for reviews see  e.g. \cite{Trev,Jrev,SYrev,Trev2}. In the simplest version they appear in the theory with 
the U$(N)$ gauge group and  $N_f=N$  quark flavors,  with the Fayet-Iliopoulos parameter (FI)
\cite{FI} $\xi\neq 0$. The non vanishing FI parameter triggers condensation of $N$ flavors of  (s)quarks, 
color-flavor locking occurs  so that both the U$(N)$ gauge group and the flavor SU$(N)$ group are broken but
the diagonal global subgroup SU$(N)_{C+F}$ survives. 

The global  SU$(N)_{C+F}$
symmetry unbroken in the vacuum but broken on the string is the reason why non-Abelian strings
(i.e. those with orientational moduli) exist. The orientational zero modes on the string solution allow 
one to rotate its color flux inside the non-Abelian SU$(N)$ group with no change in energy.
Dynamics of these orientational
moduli fields is described by an effective  two-dimensional CP$(N-1)$ model  \cite{HT1,ABEKY,SYmon,HT2}.
The emergence  of  the CP$(N-1)$ model is easy to understand. The $Z_N$ string solutions in the theory with the 
U$(N)$ gauge group break global SU$(N)_{C+F}$ down to SU$(N-1)\times U(1)$. This is why the orientational moduli live on 
\beq
CP(N-1)=\frac{ SU(N)_{C+F}}{SU(N-1)\times U(1)}.
\label{CPspace}
\eeq
The coset (\ref{CPspace}) is the target space of the sigma model on the world sheet of the simplest
Non-Abelian string. 

In the past decade many generalizations of the  simplest model were worked out.  We will continue along the lines of \cite{SYfstr} to discover a conceptual novelty: nonperturbative effects from the bulk deform the
target space of the world-sheet model (in special cases where such deformations are possible). 

In order to explain when they are possible we need to remember the following:

The CP$(N-1)$ is robust in the sense that the Ricci tensor and all higher target space covariants 
are proportional to the K\"ahler metric. The world sheet dynamics is fully characterized by a single 
parameter, the coupling constant.  In Ref. \cite{SYfstr} in which the case  $N_f>N$ and $r=N$ was 
considered  (where $r$ is the number of 
condensed quarks), we get 
the so-called weighted CP$(N,\tilde N)$ model (WCP) with $\tilde N=N_f-N$. The corresponding target space 
is not robust even in perturbation theory: extra higher target space invariants (not reducible to the 
previous) appear in higher loops. Thus, this sigma model is not renormalizable in the conventional sense. 
The $r=N$ 
vacuum considered in  \cite{SYfstr}
implies complete Higgsing of the bulk theory. There exists a limit (all quark masses large and unequal) 
in which we could derive the world-sheet WCP$(N,\tilde N)$ model that was weakly coupled.\footnote{
To be more exact, the theory on the semilocal non-Abelian 
string is the so-called  
$zn$ model \cite{SYV,KSVY}. It reduces to the WCP$(N,\tilde N)$ model at $N\to\infty$
and, at finite $N$, in the BPS-protected sector.}
Quantum 
corrections could be obtained ``inside" this  two-dimensional model {\em per se}. Then, in the BPS 
sector analytic continuation was possible to smaller masses (and mass differences).

In this paper we will consider the $r=N-1$ vacuum, with the residual unbroken U(1) in the bulk vacuum.
In this case the situation with the world-sheet model turns out dramatically different. 
At the classical level it is still WCP, but a certain corner of its target space exhibits strong coupling 
for any values of the bulk parameters. It turns out that nonperturbative corrections in the bulk 
(represented by the
gluino condensate) penetrate into the two-dimensional field theory deforming the standard 
representation for the WCP model. Correspondingly, the BPS spectrum receives nonperturbative corrections that
go beyond those occurring in the conventional WCP. 

Both theories -- that of Ref. \cite{SYfstr} and of the present work -- share a remarkable feature. 
The two-dimensional sigma models we deal with are derived as world-sheet theories on the non-Abelian strings. 
We are mainly focused on the ${\mathcal N}= (2,2)$ limit of these sigma models corresponding to
the $\mu\to 0$ limit in the bulk theory, where $\mu$ is the deformation parameter (see below). 
The ${\mathcal N}= (2,2)$ limit of the  two-dimensional sigma models of interest exists and is well-defined. 
At the same time,
in the limit of vanishing $\mu$ strings in the bulk disappear (since so do all (s)quark condensates). 
Thus, the situation we encounter with reminds the Cheshire cat's smile. The smile is there while the cat 
is gone!
The model we discuss here is even more spectacular since even at $\mu\neq 0$ one of the strings is absent.

As was already mentioned, in the
simplest model with $N_f=N$ in the $r=N$ vacuum we obtain the   CP$(N-1)$ model   on the world sheet
 of the non-Abelian string with \ntwot supersymmetry. 
Due to the (s)quark condensation in the bulk theory, the monopoles are confined. 
In the U$(N)$ theories confined elementary 
monopoles are seen as junctions of two distinct non-Abelian strings, rather then string endpoints.
They are also seen in the world-sheet theory --
as kinks interpolating between two distinct vacua of the CP$(N-1)$ model.\footnote{
Note that the \ntwot supersymmetric  CP$(N-1)$ model has $N$ vacua associated with $N$ different elementary 
strings of the bulk theory.}

This picture leads to an absolute coincidence of the BPS spectra of the bulk \ntwo QCD (in the chosen vacuum
in which $N$ (s)quark flavors  condense, i.e. $r=N$) and \ntwot supersymmetric  CP$(N-1)$ model.
This is referred to as the 2D-4D correspondence.
  This coincidence was first
observed in \cite{Dorey} and then explained using the  picture \cite{SYmon,HT2} of monopoles confined 
to non-Abelian strings.\footnote{The important 
point here is that in the simplest version of this 2D-4D correspondence both BPS spectra
do not depend on the FI parameter $\xi$ \cite{SYmon,HT2}, for a more detailed discussion see Sec. 3.
In fact, due to $\xi$ independence, the 2D-4D correspondence can be interpreted as the coincidence between the BPS spectrum of the world-sheet CP$(N-1)$ model 
and that of 
the bulk theory taken at $\xi=0$ (i.e. at a certain point on the Coulomb branch).}

In this paper we extend this 2D-4D correspondence to other vacua of 
\ntwo supersymmetric QCD, namely,  $r=N-1$. 

If the bulk ${\mathcal N}=2$ theory is perturbed by a small mass term $\mu$ for the  adjoint matter,
the Coulomb branch is lifted and the theory has the so-called $r$ vacua, where $r$ is the  number of
 condensed (s)quarks (in the large quark mass limit). The value of $r$ cannot exceed the rank of the gauge group,
i.e. $r\le N$. If all quark masses are equal this deformation does not break \ntwo supersymmetry and, in fact, 
reduces to the FI term to the leading order in $\mu$ in the $r=N$ vacuum  \cite{HSZ,VY,SYfstr}. In the  $r=N-1$ vacuum 
$N-1$ (s)quarks and no monopoles condense. The absence of the condensed monopoles singles out
 $r=N-1$.
Below we assume that quark masses
 are generic so all $r$-vacua are isolated, no Higgs branches appear.

First, we obtain the  classical theory on the non-Abelian string in the $r=N-1$ vacuum. This is quite easy 
since it  is given by a 
WCP model.
Then we find a quantum deformation of the model
using the method of resolvents suggested recently by Gaiotto, Gukov 
and Seiberg for surface defects \cite{GGS}.\footnote{Certain surface defects are related to  
non-Abelian strings in the low-energy limit. In our language the Gaiotto-Gukov-Seiberg setup  can
 be understood as  gauging of the flavor group and sending $\xi\to\infty$ \cite{Gaiotto}.
In this limit all massive bulk states decouple, and  non-Abelian strings  become infinitely thin and infinitely heavy.
In this paper we do not gauge the flavor group and consider  finite values of $\xi$.}
 In much the same way as 
in the $r=N$ vacuum \cite{SYfstr}, the bulk monopoles are seen as kinks in the world-sheet theory. 
In the $\mu\to 0$ limit all vacua of the world-sheet theory are degenerate, and kinks are static.
(If $\mu\neq 0$ the degeneracy is lifted, and strictly speaking there are no static kink solutions.)

Our calculation demonstrates the 
coincidence of 2D and 4D BPS spectra. The 2D-4D correspondence holds in the $r=N-1$ vacua.

The paper is organized as follows. In Sec.~\ref{bulk} we briefly outline the structure of the $r$ vacua in the $\mu$-deformed \ntwo
QCD. In Sec.~\ref{r=Nvacuum} we review the world sheet-theory  and 2D-4D correspondence 
in these  vacua. These two sections, Secs. 2 and 3, are needed to introduce relevant notation
and specify our overall setting. 
Then we proceed to new results in the  $r=N-1$ vacuum. In Sec.~\ref{r=N-1vaccl} the effective theory on 
the non-Abelian string in the $r=N-1$ vacuum is considered at the classical level. 
In Sec. \ref{quantum} we study its quantum deformation. In Sec. \ref{2D/4D} we prove the coincidence 
of the BPS spectra  in
the world-sheet and bulk theories. We then calculate the kink 
mass in the semiclassical approximation in the simplest $N=2$ case. In Sec. \ref{deformation} we discuss
the $\mu$-dependent deformation potential in the world-sheet theory while 
Sec. \ref{concl} summarizes our conclusions. In Appendices A and B we present semiclassical
calculations of the roots of the Seiberg-Witten curve and the monopole mass in the U(2) bulk theory,
 respectively.

\section {\boldmath{$r$} Vacua in \ntwo QCD}
\label{bulk}
\setcounter{equation}{0}

\subsection{ \boldmath{$\mu$}-Deformed \boldmath{\ntwo} QCD}
\label{model}

The gauge symmetry of our basic model is 
U($N$)=SU$(N)\times$U(1). In the absence
of  deformation the model under consideration is \ntwo  SQCD
 with $N_f$ massive quark hypermultiplets. 
 We assume that
$N_f \ge N$ but $N_f< 2 N$. 
The latter inequality ensures  the  theory to be asymptotically free. 

In addition, we will introduce the mass term $\mu$ 
for the adjoint matter breaking \ntwo supersymmetry down to \none\!. 
Thus, the bulk theory is essentially the same as in \cite{SYfstr}.
The \ntwo vector multiplet
consists of the  U(1)
gauge field $A_{\mu}$ and the SU$(N)$  gauge field $A^a_{\mu}$,
where $a=1,..., N^2-1$, and their Weyl fermion superpartners plus
complex scalar fields $a$, and $a^a$ and their Weyl superpartners, respectively.
The $N_f$ quark multiplets of  the U$(N)$ theory consist
of   the complex scalar fields
$q^{kA}$ and $\tilde{q}_{Ak}$ (squarks) and
their   fermion superpartners --- all in the fundamental representation of 
the SU$(N)$ gauge group.
Here $k=1,..., N$ is the color index
while $A$ is the flavor index, $A=1,..., N_f$. We will treat $q^{kA}$ and $\tilde{q}_{Ak}$
as rectangular matrices with $N$ rows and $N_f$ columns. 

Let us first discuss the undeformed  \ntwo theory.
 The  superpotential is
 \beq
{\mathcal W}_{{\mathcal N}=2} = \sqrt{2}\,\sum_{A=1}^{N_f}
\left( \frac{1}{ 2}\,\tilde q_A {\mathcal A}
q^A +  \tilde q_A {\mathcal A}^a\,T^a  q^A + m_A\,\tilde q_A q^A\right)\,,
\label{superpot}
\eeq
where ${\mathcal A}$ and ${\mathcal A}^a$ are  chiral superfields, the ${\mathcal N}=2$
superpartners of the gauge bosons of  U(1) and SU($N$), respectively.
Then we add a mass term for the adjoint fields 
\beq
{\mathcal W}_{{\rm def}}=
  \mu\,{\rm Tr}\,\Phi^2, \qquad \Phi\equiv\frac12\, {\mathcal A} + T^a\, {\mathcal A}^a
\label{msuperpotbr}
\eeq
which breaks supersymmetry down to ${\mathcal N}=1$, generally speaking.
However, to the leading order in  $\mu$
   and if all quark masses are equal this term reduces to the 
Fayet-Iliopoulos $F$ term  which can be rotated  into the $D$ term \cite{FI}. The latter  
does not break \ntwo supersymmetry  \cite{HSZ,VY,SYmon,SYfstr}.

\subsection{\boldmath{$r$} Vacua}
\label{rv}

The $r$ vacuum is a vacuum with $r$ flavors of (s)quarks condensed.
The $r$ counting is assumed to be carried out  in the weak coupling domain
 at large quark masses.  It is obvious that
the maximal value of  $r$ is $N$. The number of the isolated $r=N$ vacua is  
\beq
{\cal N}_{r=N} = C_{N_f}^{N}= \frac{N_f!}{N!(N_f-N)!}\,,
\label{numNvac}
\eeq
 see \cite{SYrev}. All gauge   bosons are completely Higgsed, and the
theory is in the color-flavor locked phase (assuming the quark masses to be close to each other). 
The (s)quark vacuum expectation values (VEVs) are determined by
\beq
\xi_P \sim  \mu m_P\,,\qquad P=1,..., N\,.
\eeq
A more precise definition of the set of the parameters $\xi^P$ will be given below, see Eqs. (\ref{qvevr}),
 (\ref{xiclass}) and (\ref{xirN}).
  For large values of $\xi$ the bulk theory is at weak coupling. Then
it can be studied semiclassically. In particular, non-Abelian strings  confining monopoles are known to exist
\cite{HT1,ABEKY,SYmon,HT2}.

For generic $m_A$    
the number of the isolated $r$ vacua with $r<N$  is \cite{CKM}
\beq
{\cal N}_{r<N}=\sum_{r=0}^{N-1} \,(N-r)\,C_{N_f}^{r}= \sum_{r=0}^{N-1}\, (N-r)\,\frac{N_f!}{r!(N_f-r)!}\,
\label{nurvac}
\eeq
representing the number of choices one can pick up $r$ condensing quarks out of $N_f$ quarks times the
Witten index  in the classically unbroken SU$(N-r)$ pure gauge theory. 

Consider a
 particular vacuum in which the first $r$ quarks develop nonvanishing VEVs. 
 Quasiclassically, at large masses, the adjoint scalar VEVs   are  
\beq
\Phi^{\rm cl} = - \frac1{\sqrt{2}}\,
{\rm diag}\left[m_1,...,m_r,0, ..., 0 
\right].
\label{avevr}
\eeq
The last $(N-r)$ entries vanish at the   classical level. For those quarks which condense the 
corresponding eigenvalue of $\Phi$ is determined by the mass of this quark, while for those quarks which 
do not condense the corresponding eigenvalue should be a critical point of the deformation superpotential
(\ref{msuperpotbr}), which vanishes.

At the quantum level these zero entries acquire values determined by $\Lambda$, 
where $\Lambda$ is the scale of \ntwo QCD.
In the classically unbroken U$(N-r)$ pure gauge sector the gauge symmetry gets broken through the 
Seiberg--Witten mechanism \cite{SW1,SW2}:
first down to U(1)$^{N-r}$ and then almost completely by condensation of $(N-r-1)$ monopoles. A single
 U(1) gauge factor survives.

This unbroken U(1) factor
in all $r<N$ vacua makes them critically different from the $r=N$ vacuum: in the latter there are
no long-range forces. 

Consider the non-Abelian limit when quark mass differences $\Delta m_{AB}=m_A-m_B$ are small,
$\Delta m_{AB}\ll m_A$.
The low-energy theory in the $r$ vacuum  has the gauge group
\beq
{\rm U}(r)\times {\rm U}(1)^{N-r}\,,
\label{legaugegroup}
\eeq
 with $N_f$ quark flavors  charged
under the U$(r)$ factor and $(N-r-1)$ monopoles charged under the U(1) factors.

For $r> N_f/2$ and large $\xi \sim \mu m$ the SU$(r)$ non-Abelian
quark sector is at weak coupling since it is asymptotically free.\footnote{The opposite case $r< N_f/2$
is discussed in \cite{SYhybrid}.}
The  quark condensates can be read-off from the superpotentials (\ref{superpot}) 
and (\ref{msuperpotbr}) using 
(\ref{avevr}). They are 
\beqn
\langle q^{kA}\rangle &=& \langle\bar{\tilde{q}}^{kA}\rangle=\frac1{\sqrt{2}}\,
\left(
\begin{array}{cccccc}
\sqrt{\xi_1} & \ldots & 0 & 0 & \ldots & 0\\
\ldots & \ldots & \ldots  & \ldots & \ldots & \ldots\\
0 & \ldots & \sqrt{\xi_r} & 0 & \ldots & 0\\
\end{array}
\right),
\nonumber\\[4mm]
k&=&1,..., r\,,\qquad A=1,...,N_f\, ,
\label{qvevr}
\eeqn
with all other components vanishing.
The first  $r$ parameters $\xi$ in the  quasiclassical  approximation are 
\beq
\xi_P \approx 2\;\mu m_P,
\qquad P=1,..., r\,.
\label{xiclass}
\eeq
These parameters can be made large in the large $m_A$ limit even if  $\mu$ is  small.

\subsection{Quantum effects}
\label{qef}

In quantum theory all parameters $\xi_P$  are determined by the roots of the Seiberg-Witten (SW) curve
\cite{SYfstr,SYrvacua,SYhybrid} which in the case at hand 
 takes the form \cite{APS}
\beq
y^2= \prod_{P=1}^{N} (x-\phi_P)^2 -
4\left(\frac{\Lambda_{{\mathcal N}=2}}{\sqrt{2}}\right)^{2N-N_f}
\, \,\,\prod_{A=1}^{N_f} \left(x+\frac{m_A}{\sqrt{2}}\right).
\label{curve}
\eeq
Here $\phi_P$ are gauge invariant parameters on the Coulomb branch. Instead of (\ref{avevr}) one must write
\beq
\Phi \approx 
{\rm diag}\left[\phi_1,...,\phi_N\right],
\eeq
where
\beq
\phi_P \approx -\frac{m_P}{\sqrt{2}},\quad P=1, ... ,\, r\,; \qquad 
\phi_P \sim \Lambda,\quad P=r+1, ... ,\, N\,.
\label{classphi}
\eeq

In the $r=N$ vacuum the curve (\ref{curve}) has $N$ double roots associated with condensation
 of $r=N$ quarks and reduces to
\beq
y^2= \prod_{P=1}^{N} (x-e_P)^2,
\label{rNcurve}
\eeq
where quasiclassically (at large masses) $e_P$'s are given  by the
mass parameters, 
$$\sqrt{2}e_P\approx -m_P\,,\qquad  P=1, ... , N\,.
$$
In fact, the (s)quark condensates  in the $r=N$ vacuum are determined by the  exact formula \cite{SYfstr}
\beq
\xi_P=-2\sqrt{2}\,\mu\,e_P\,.
\label{xirN}
\eeq

Now, consider the $r<N$ vacua. As was mentioned, in
this paper we will focus on the simplest example of the $r<N$ vacuum, namely, $r=N-1$. 
In this vacuum $r=N-1$ quarks condense in the large $m_A$ limit, and there are no light
(condensed) monopoles.
To identify the $r=N-1$ vacuum in terms of the curve (\ref{curve}) it is necessary to find
such values of $\phi_P$ which ensure the Seiberg-Witten curve to have $N-1$ double roots with 
 $\phi_P$   approximately determined by the quark masses, see (\ref{classphi}).

We know that the Seiberg--Witten curve factorizes \cite{CaInVa},
\beq
y^2
=\prod_{P=1}^{N-1} (x-e_P)^2\,(x-e_N^{+})(x-e_N^{-})\,.
\label{rcurve}
\eeq
The first $r=N-1$ double roots are determined by the
mass parameters in the large mass limit, $\sqrt{2}e_P\approx  -m_P$, $P=1, ... , (N-1)$. 
The last two  roots  are  much smaller and are determined by $\Lambda$. 
 For the single-trace deformation superpotential (\ref{msuperpotbr}) 
 their sum vanishes \cite{CaInVa},
\beq
e_N^{+} + e_N^{-}=0\,.
\label{DijVafa}
\eeq
The root $e_N^{+}$ determines the value of the gaugino condensate \cite{Cachazo2},
\beq
\left( e^\pm_N\right)^2=\frac{2S}{\mu}, \qquad S=\frac1{32\pi^2}\langle {\rm Tr}\,W_{\alpha}W^{\alpha} \rangle,
\label{eN}
\eeq
where the superfield $W_{\alpha}$ includes the gauge field strength tensor.

In terms of the roots of the Seiberg-Witten curve the quark VEVs  are given by 
the formula\,\footnote{In fact Eq.~(\ref{xi}) is  very general and determines the condensed state
VEVs, namely, quarks or monopoles, independently of their nature in any vacuum with $r<N$
\cite{SYhybrid}. In the particular  $r=N-1$ vacuum which  we consider here there are no condensed monopoles.} \cite{SYrvacua,SYhybrid}
\beq
\xi_P=-2\sqrt{2}\,\mu\,\sqrt{(e_P-e_N^{+})(e_P-e_N^{-})}
\label{xi}
\eeq
for $P=1,...,(N-1)$.

\subsection{Non-Abelian strings}
\label{nastrings}

As was already mentioned, our  theory   supports non-Abelian  strings
 \cite{HT1,ABEKY,SYmon,HT2}. At  $\mu\ll (m_A, \Lambda )$ these strings are BPS-saturated  \cite{HSZ,VY} 
and their tensions 
 are determined exactly  by the  $\xi$ parameters, namely  \cite{SYrev,SYfstr}
 \beq
T_P=2\pi|\xi_P|,
\label{ten}
\eeq
with $\xi_P$ given by  (\ref{xirN}) and (\ref{xi})   in  the $r=N$ and $r<N$ vacua, respectively.
 
Both in the $r=N$ and $r=N-1$ vacua non-Abelian strings are magnetic and confine monopoles.
More precisely, the elementary monopole $M_{P,P+1}$ at   $\mu\neq 0$ becomes the junction of $P$-th and
$(P+1)$-th elementary non-Abelian strings, $P=1,...(N-1)$.

\begin{figure}
\epsfxsize=8cm
\centerline{\epsfbox{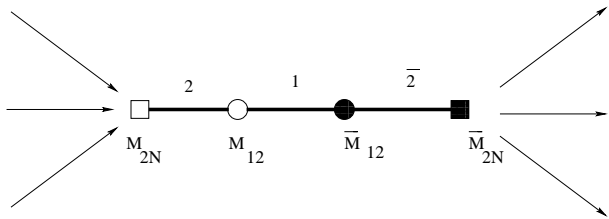}}
\caption{\small Example of the dipole meson formed as a result 
of breaking of the ``second" string by pair creation of the monopole
$M_{2N}$ (shown by boxes) interpolating between the ``second" string and the would-be $N$-th string
(which is absent).
Arrows denote unconfined flux.
Circles denote monopoles $M_{PP'}$, $P,P'=1,...,(N-1)$. 
Open and closed circles/boxes denote  monopoles and antimonopoles, respectively.}
\label{figdipole}
\end{figure}

In the $r=N-1$ vacuum there is a peculiar feature distinguishing it from the $r=N$ vacuum. One string (say, the 
$N$-th
string) is absent because the $N$-th quark does not condense. 
As a result, all strings become metastable. They can be broken by a pair creation of particular monopoles which 
interpolate between the $P$-th string  and 
the  would-be $N$-th string, which is in fact absent ($P=1,...,(N-1)$). 
An example of the monopole meson emerging in this way
is shown in Fig.~\ref{figdipole}. 

The endpoints emit fluxes of the unbroken U(1) gauge field. 
This makes this meson a dipole-like configuration. 
Note that the non-Abelian fluxes of the SU$(N-1)$ gauge group are always trapped and squeezed
in the non-Abelian strings.
Long-range forces are associated only with the unbroken U(1) gauge factor.

In the large mass limit the masses of the monopoles which break the string become large, of the order of 
$m/g^2$ (where $g^2$ is the coupling constant in the SU($N$) group), and strings become metastable.

\section {2D-4D correspondence in the \boldmath{$r=N$} \\
vacuum}
\label{r=Nvacuum}
\setcounter{equation}{0}

In this section we will briefly review non-Abelian strings in the $r=N$ vacuum and associated 2D-4D correspondence.
We will start with the simplest version of the bulk theory with the  FI $D$-term and then pass to $\mu$-deformed \ntwo QCD.

\subsection{Bulk theory with the FI term }
\label{FIterm}

Consider \ntwo QCD with the FI term of the $D$ type. For simplicity we will assume now that $N_f=N$. As was already
 mentioned, the dynamics of orientational zero modes of non-Abelian string which become orientational moduli fields 
 on the world sheet is described by two-dimensional
\ntwot supersymmetric CP$(N-1)$ model, see e.g. \cite{SYrev} for a review.
 This model can be nicely written as a U(1) gauge theory in the strong coupling limit (the so-called gauged formulation) \cite{W79}. 
The bosonic part of the action is
\beqn
S_{{\rm CP}(N-1)}
& =&
\int d^2 x \left\{
\left|\nabla_{\alpha} n^{P}\right|^2 +\frac1{4e^2}F^2_{\alpha\beta} + \frac1{e^2}
|\pt_\alpha\sigma|^2
\right.
\nonumber\\[3mm]
 &+&  
\left|\sigma+ m_P\right|^2 |n^{P}|^2 + 
\frac{e^2}{2} \left(|n^{P}|^2 -2\beta\right)^2
\Big\}\,,
\label{cpg}
\eeqn
where $n^P$ are complex fields, $P=1,..., N$, $$\nabla_{\alpha}= \partial_{\alpha} - i A_{\alpha} \,,$$ 
 $\sigma$ is a complex scalar
field, and summation over $P$ is implied. The condition 
\beq
  |n^P|^2 =2\beta
\label{unitvec}
\eeq 
is implemented in the limit $e^2\to\infty$. Moreover, in this limit
the gauge field $A_{\alpha}$  and its \ntwo bosonic superpartner $\sigma$ become
auxiliary and can be eliminated by virtue of the equations of motion.

In the limit of equal quark masses the global SU($N)_{C+F}$ symmetry is unbroken, and 
strings are fully non-Abelian. This is a strong coupling quantum regime
in the CP$(N-1)$ model (\ref{cpg}). The vector $n^P$ is smeared all over the entire CP$(N-1)$
space due to quantum fluctuations and its average value vanishes \cite{W79}.
 The world-sheet theory develops a mass gap $\Lambda \ll \sqrt \xi$.

At small nonvanishing $\left|m_P-m_{P'}\right|$ the global SU($N)_{C+F}$ symmetry is 
explicitly broken down to $U(1)^{(N-1)}$.
A shallow potential is generated on the  CP$(N-1)$ moduli space as is seen from
(\ref{cpg}). As we increase $\left|m_P-m_{P'}\right|$ the strings become ``more Abelian"
and eventually evolve into Abelian $Z_N$ strings, which correspond to
$N$ classical vacua of the world-sheet model (\ref{cpg})
\beq
n^P=\sqrt{2\beta}\;\delta^{PP_0},\qquad \sigma=-\,m_{P_0},
\label{cphiggsvac}
\eeq
where $P_0$ can take any of $N$ values, $P_0=1 , ..., N$, see the review \cite{SYrev}.

The two-dimensional coupling constant $\beta$  ($\beta =1/g^2$) is determined by the
four-dimensional non-Abelian coupling $g$ via the relation
\beq
\beta= \frac{2\pi}{g^2_{ 4D}}\,.
\label{betag}
\eeq
This relation is valid at the inverse transverse size of the string given by $g\sqrt{\xi}$ 
 which plays the role of ultra-violet cutoff of the effective
theory (\ref{cpg}) on the string, see the review \cite{SYrev}. Given that $\beta$-functions 
of the bulk and world-sheet theories are the same this leads to the following identification
\beq
\Lambda_{2D}=\Lambda_{4D},
\label{equallambdas}
\eeq
which plays an important role in the coincidence of the BPS spectra of two theories.

\subsection{More flavors}

Adding ``extra" quark flavors with degenerate masses we increase $N_f$ from $N$ up to a certain value
$N_f >N$.
The strings emerging in such theory are semilocal.
In particular, the string solutions on the Higgs branches (typical
for multiflavor theories) usually are not fixed-radius strings, but, rather,
possess radial moduli, also known as the size moduli, see   \cite{AchVas} for a comprehensive review of 
the Abelian semilocal strings. The transverse size of such strings is not fixed.

Non-Abelian semilocal strings in \ntwo SQCD with $N_f>N$ were studied in
\cite{HT1,HT2,SYsem,Jsem,SYV}. 
The orientational
moduli of the semilocal non-Abelian string can be described by a complex 
vector $n^P$ (here $P=1, ..., N$),
 while its $\tN=(N_f-N)$ size moduli are parametrized by a complex vector
$\rho^K$ ($K=N+1, ..., N_f$). The effective two-dimensional theory
which describes the internal dynamics of the non-Abelian semilocal string is
the \ntwot weighted  CP model, which includes both types of fields. The bosonic 
part of the action
in the gauged formulation (which assumes taking the limit $e^2\to\infty$)
has the form
\beqn
&&S_{\rm WCP} = \int d^2 x \left\{
 \left|\nabla_{\alpha} n^{P}\right|^2 
 +\left|\tilde{\nabla}_{\alpha} \rho^K\right|^2
 +\frac1{4e^2}F^2_{\alpha\beta} + \frac1{e^2}\,
\left|\pt_{\alpha}\sigma\right|^2
\right.
\nonumber\\[3mm]
&+&\left.
\left|\sigma+ m_P\right|^2 \left|n^{P}\right|^2 
+ \left|\sigma+ m_{K}\right|^2\left|\rho^K\right|^2
+ \frac{e^2}{2} \left(|n^{P}|^2-|\rho^K|^2 -2\beta\right)^2
\right\},
\nonumber\\[4mm]
&& 
P=1,..., N\,,\qquad K=N+1,..., N_f\,.
\label{wcp}
\eeqn
The fields $n^{P}$ and $\rho^K$ have
charges  +1 and $-1$ with respect to the auxiliary U(1) gauge field;\footnote{In fact,
 the theory on the semilocal non-Abelian 
string is not exactly the WCP model (\ref{wcp}). The actual theory that emerges on the world sheet was called the
$zn$ model \cite{SYV,KSVY}. It has a somewhat different metric of the 
target space. The WCP model (\ref{wcp}) correctly reproduce the BPS spectrum of the 
world-sheet theory, which comes out exactly the same as in the
$zn$ model \cite{SYV,KSVY}. In what follows we use a similar approach  to describe the BPS spectrum of
the 2D theory on the non-Abelian string in the $r=N-1$ vacuum.} 
hence, the corresponding  covariant derivative $\tilde{\nabla}$  in (\ref{wcp}) is$$\tilde{\nabla}_{\alpha}=\d_{\alpha}+iA_{\alpha}\,.$$

As in the CP$(N-1)$ model, small mass differences
$\left| m_A-m_B\right|$ lift orientational and size zero modes generating a shallow potential on the
 modular space.

The coupling constant $\beta$ in (\ref{wcp}) is related to the bulk coupling via (\ref{betag})
which ensures the coincidence of scales of bulk and world sheet theory, see (\ref{equallambdas}),
where we used  that the first coefficient of the $\beta$ function $b=2N-N_f$ is the same for the bulk and 
world-sheet theories.

\subsection{Exact superpotential}
\label{es}

An  exact twisted superpotential of the  Veneziano-Yankielowicz  type \cite{VYan} is known  in
the \cpn model \cite{AdDVecSal,ChVa,W93,Dorey}.
This superpotential was later generalized to the case of the WCP models in 
\cite{HaHo,DoHoTo}.
Integrating out the fields $n^P$ and $\rho^K$  we obtain
 the following
exact twisted superpotential:
\beqn
 &&{\cal W}_{\rm WCP}(\sigma)= 
\frac1{4\pi}\left\{\sum_{P=1}^N\,
\left(\sigma+{m}_P\right)
\,\ln{\frac{\sigma+{m}_P}{\Lambda}}
\right.
\nonumber\\[3mm]
&& 
\left.
-\sum_{K=N+1}^{N_F}\,
\left(\sigma+{m}_K\right)
\,\ln{\frac{\sigma+{m}_K}{\Lambda}} 
- (N-\tN) \,\sigma \right\}\, ,
\label{CPsup}
\eeqn
where we use one and the same notation $\sigma$ for the  twisted superfield \cite{W93} and its lowest scalar
component. 
Minimizing this superpotential with 
respect to $\sigma$ we get the equation for the VEVs of $\sigma$ (the so-called twisted chiral ring equation),
\beq
\prod_{P=1}^N(\sigma+{m}_P)
=\Lambda^{(N-\tN)}\,\prod_{K=N+1}^{N_f}(\sigma+{m}_K)\,.
\label{sigmaeq}
\eeq

The  masses of the BPS kinks interpolating between the
vacua $\sigma_{P}$ and $\sigma_{P'}$ are given  by the appropriate 
differences of the superpotential (\ref{CPsup}) calculated at distinct roots \cite{HaHo,Dorey,DoHoTo},
\beq
M^{\rm BPS}_{PP'} =
2\left|{\cal W}_{\rm WCP}(\sigma_{P'})-{\cal W}_{\rm WCP}(\sigma_{P})\right|\,,\qquad P,P'=1,..., \,N\,.
\label{BPSmass}
\eeq
Due to the presence of branches  in the logarithmic functions in (\ref{CPsup}) each kink come together with a 
tower of dyonic kinks carrying global U(1) charges (for more details
see e.g. \cite{Bolokhov:2012dv}). In addition to  kinks the BPS spectrum of the model
contains elementary excitations with masses given by $|m_A-m_P|$, $A=1,...,N_f$,  $P=1,...,N$.

The  masses obtained from (\ref{BPSmass}) were shown  
to coincide with those of the monopoles and dyons in the bulk theory. The latter are 
given by the period integrals of the Seiberg--Witten curve (\ref{curve}). 

 As was mentioned in Sec.~\ref{intro},
this coincidence was observed in \cite{Dorey,DoHoTo} and   explained later 
in \cite{SYmon,HT2} using the picture of confined bulk monopoles which are seen as kinks in the world 
sheet theory. A crucial point is that both monopoles and kinks are BPS-saturated states\,\footnote{Confined
 monopoles, being junctions of two distinct 1/2-BPS strings, are 1/4-BPS states in the bulk theory 
\cite{SYmon}.},
and their masses cannot depend on the non-holomorphic parameter $\xi$ \cite{SYmon,HT2}. This means that,
although confined monopoles look physically very different from unconfined monopoles on the Coulomb branch
of the bulk theory (in the particular singular point which becomes the $r=N$ vacuum at nonzero $\xi$),
their masses are the same. Moreover, they coincide with the masses of kinks in the world-sheet 
theory.

Note that the roots of the vacuum  equation (\ref{sigmaeq}) coincide with the double roots of the 
Seiberg--Witten curve (\ref{curve}) of 
the bulk theory \cite{Dorey,DoHoTo},
\beq
\sigma_P=\sqrt{2}\,e_P\, .
\label{equalroots}
\eeq
This is the key technical reason which leads to the coincidence of the BPS spectra.

\subsection{The \boldmath{$r=N$} vacuum in \boldmath{$\mu$}-deformed \boldmath{${\mathcal N}=2$}  QCD}

\begin{figure}[h]
\epsfxsize=10cm
\centerline{\epsfbox{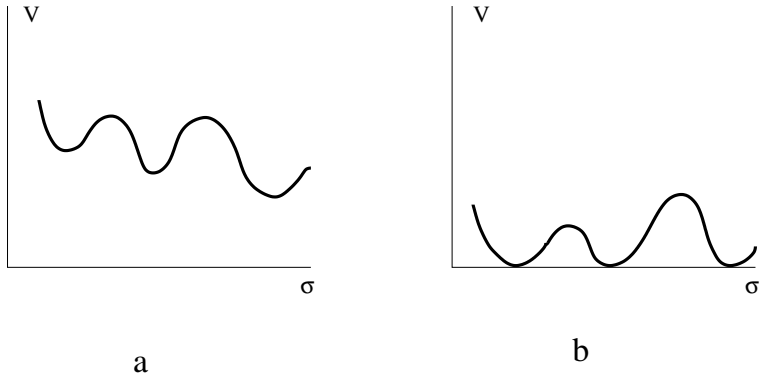}}
\caption{\small a. Schematic picture of the scalar potential in the theory (\ref{S2Dr=N}).
The complex variable $\sigma$ is schematically represented by the horizontal axis. Minima
of the potential correspond to elementary non-Abelian strings with tensions given by (\ref{Vten}).
b. The same potential in the limit $\mu=0$.
}
\label{r=Npot}
\end{figure}

Now let us switch off the Fayet-Iliopoulos $D$  term in the bulk theory and consider instead 
the $F$ term deformation (\ref{msuperpotbr}).
In \cite{SYfstr,BSYadj} it was shown that at generic quark masses \ntwot supersymmetry is broken down to
\ntwoo  even to the leading order in $\mu$. For the single-trace deformation  (\ref{msuperpotbr})
the bosonic part of the low-energy world-sheet theory becomes
\beq
S_{2D}=S_{(2,2)}+ \int d^2 x\, V_{\rm def} (\sigma),
\label{S2Dr=N}
\eeq
where $S_{(2,2)}$ is the action of \ntwot supersymmetric model (\ref{wcp}) while 
the deformation potential is given by
\beq
V_{\rm def} (\sigma)=4\pi\,|\mu\sigma|\,.
\label{Vdefr=N}
\eeq
The total scalar potential given by the sum of the twisted mass potential in (\ref{wcp}) and deformation 
(\ref{Vdefr=N}) is schematically shown in Fig.~\ref{r=Npot}\,a. Its $N$ minima correspond to tensions of 
$N$ elementary non-Abelian strings,
\beq
V(\sigma_P) = T_P, \qquad P=1,...,N.
\label{Vten}
\eeq
To see this we note that the vacuum values $\sigma_P$ are still given by solutions of the chiral ring equation
(\ref{sigmaeq}) corresponding to  the limit  $\mu\to 0$. Then the
coincidence of the roots of the bulk and world-sheet theory
(\ref{equalroots}), together with Eqs.~(\ref{xirN}) and (\ref{ten}), gives (\ref{Vten}).
At $\mu\neq 0$ and generic masses the minima are non-degenerate, only the lowest lying vacuum is stable,
no static kink solutions exist.

The stability of the lowest vacuum in two dimensions means that
the lightest  of $N$ non-Abelian strings is stable, others become metastable. Moreover, since
generically string tensions do not vanish, \ntwoo supersymmetry is broken spontaneously already at the classical
level \cite{SYfstr}. The barriers between different vacua of the potential in Fig.~\ref{r=Npot}
are of the order of the quark mass differences $|m_A-m_B|^2$.

The 2D-4D correspondence manifests itself as follows. Both the confined  bulk monopoles
and kinks of the world-sheet theory are no longer BPS-saturated, and we cannot expect that their masses
are independent of $\mu$ (or, which is the same, $\xi$ in the case at hand). We expect that the kink mass is   
\beq
M^{\rm kink}_{PP'} =
2\left|{\cal W}(\sigma_{P'})-{\cal W}(\sigma_{P})\right|+ O(\mu)\,,
\qquad P,P'=1,...,\,N\,,
\label{kinkmassgen}
\eeq
where ${\cal W}(\sigma)= {\cal W}_{\rm WCP}$  for the $r=N$ vacuum (see (\ref{CPsup})) and the term 
$O(\mu)$ represents  non-BPS $\mu$-corrections. Strictly speaking, the kink is not defined as a static object at
$\mu\neq 0$. Because of the difference of tensions it must accelerate.

The mass of the confined monopole is given by 
\beq
M^{\rm monopole}_{PP'}=\left|\frac{\sqrt{2}}{2\pi i}\,\oint_{\beta_{PP'}} d\lambda_{SW}\right|
 +O(\mu)\,,
\qquad P,P'=1,....N,
\label{monmassgen}
\eeq
where integral of the Seiberg-Witten differential \cite{SW1,SW2,ArFa,KLTY,ArPlSh,HaOz} 
goes along the $\beta$ contour through shrinking cuts associated with  double roots $e_P$ and $e_{P'}$.
 The second term represents non-BPS $\mu$-corrections.

Since the kink in the low-energy theory on the non-Abelian string represents a confined bulk monopole   
their masses should be the same,
\beq
M^{\rm monopole}_{PP'}=M^{\rm kink}_{PP'}\,, \qquad P,P'=1,...,\,N\, ,\qquad \mu\to 0\,.
\label{equalmasses}
\eeq
In other words,  the
BPS spectra of the  theory with the action $S_{(2,2)}$ and the 
bulk theory on the Coulomb branch
(at the particular singular point which becomes the $r=N$ vacuum upon $\mu$-deformation) should coincide
with each other. As was already explained, this conclusion was explicitly checked in \cite{Dorey,DoHoTo}
for the $r=N$ vacuum.  In this form the 2D-4D correspondence is easy to generalize to other $r$ vacua.
We will use this form of the 2D-4D correspondence in what follows.

It seems somewhat confusing to take limit $\mu\to 0$ in the world-sheet theory because in this limit
strings disappear while monopoles become unconfined. However,  if we forget that this  theory represents the theory on the
string and view it as a 2D theory {\em per se},  we see from Eq.~(\ref{S2Dr=N}) that
 this limit is perfectly well defined. Moreover, both kinks and monopoles become
BPS saturated in this limit. The potential of the 2D theory in this limit is shown in Fig.~\ref{r=Npot}\,b.

\section {Classical theory on the non-Abelian string in the \boldmath{$r=N-1$} vacuum}
\label{r=N-1vaccl}
\setcounter{equation}{0}

Now we start construction of the world-sheet theory in the $r=N-1$ vacuum at the classical level.

In much  the same way as in the $r=N$ vacuum the low-energy theory on the string in the 
$r=N-1$ vacuum is given by
the sum of an \ntwot supersymmetric  theory (let us call it  $T_{(2,2)}$) and a $\mu$-deformation,
\beq
S_{2D}=S_{(2,2)}+ \int d^2 x\, V_{\rm def} (\sigma),
\label{S2Dr=N-1}
\eeq
where $S_{(2,2)}$ is the action of  the $T_{(2,2)}$ theory. The deformation potential $ V_{\rm def}$
at its minima 
gives the tensions of the non-Abelian
 strings, cf. (\ref{Vdefr=N}). The string tensions  in the $r=N-1$ vacua are given by
\beq
 V_{\rm def} (\sigma_P)=T_P =4\pi\sqrt{2}\,\left|\mu\,\sqrt{(e_P-e_N^{+})(e_P-e_N^{-})}\right|
\label{vacua}
\eeq
for $P=1,..., (N-1)$, see Eqs. (\ref{xi}) and (\ref{ten}). 

Now we have only $(N-1)$ strings, while the 
$N$-th string is absent. The associated minimum of  $V_{\rm def}$ is the ground state at zero energy.

Let us find \ntwot supersymmetric theory $T_{(2,2)}$ neglecting for a while the deformation potential
 $ V_{\rm def}$ in (\ref{S2Dr=N-1}). We will discuss it later in Sec.~\ref{deformation}.
We also assume for simplicity that $N_f=N$.  

Consider first the 
quasicalssical limit
\beq
m_A\gg\Lambda, \qquad \Delta m_{AB}=(m_A-m_B)\ll m_A\,.
\label{cllimit}
\eeq
In this limit the low-energy gauge group of the bulk theory becomes
\beq
U(N-1)\times U(1),
\label{legroup}
\eeq
where the U(1) factor is unbroken, and the theory has $N_f=N$ quarks charged under the U$(N-1)$ factor, see
Eqs. (\ref{avevr}) and (\ref{legaugegroup}).
The (s)quark fields develop VEVs given by Eq. (\ref{qvevr}) with $r=N-1$.

Thus, this low-energy theory supports non-Abelian strings. Since the number of the quark flavors $N$ is 
lager than the  rank $r=N-1$ of the low-energy gauge group  by 1,  these strings are semilocal.
The world-sheet theory is given by the WCP model\,\footnote{See footnote 7.} (\ref{wcp}) with $N-1$
orientational moduli  $n^P$  with charge +1 ($P=1,..., N-1)$, plus a single size modulus $\rho$, with charge $-1$.
With the mass parameters chosen according to (\ref{cllimit}) $N-1$ strings are (meta)stable. 

The coefficient $b$ of the $\beta$ function of this
 low-energy world-sheet theory is  the sum of charges of the $n^P$ and $\rho$ fields, namely
$b_{LE}= (N-1)-1 = N-2$. This coefficient  coincides with the coefficient $b$
of the bulk theory in the low-energy limit, $b_{LE}= 2(N-1)-N_f = N-2$.

Now,  let us relax the
condition  $\Delta m_{AB}\ll m_A$. To determine the world-sheet theory $T_{(2,2)}$ we can use the 
following procedure. 

Let us start from the $r=N$ vacuum where the theory on the non-Abelian string is given by the
CP$(N-1)$ model (\ref{cpg}),  with the $\beta$-function coefficient   $b=N$. Then we  reduce the mass of the 
$N$-th quark $m_N$.
The point $m_N=0$ is a point where two vacua ($r=N$ and $r=N-1$) coalesce.
At this point we can ``jump'' into the $r=N-1$ vacuum and then increase 
$m_N$ to its initial value. In this process our world-sheet theory
is  smoothly deformed from the CP$(N-1)$ model to the theory $T_{(2,2)}$ sought for. 

This implies that the 
theory $T_{(2,2)}$ is given by the CP$(N-1)$ model in which the ``last" field $n^N$ is taken with $m_N=0$
plus an extra conformal sector which does not spoil the correct $\beta$ function.
 The point is  that the coefficient $b$ of the $\beta$ function of 
 world-sheet theory should coincide with the one for the bulk theory, $b=N$.

Thus, the conformal sector must consist of two complex fields $z$ and 
$\rho$, with charges $+1$ and $-1$, respectively. At large masses in the 
limit (\ref{cllimit}) the $n^N$ field  present in the CP$(N-1)$ model, as well as 
$z$, become massive and decouple, so we are left with the  low-energy WCP model
described after Eq. (\ref{legroup}), see Eq. (\ref{Tcl}) with $n^N$ and $z$ crossed out.

Combined with \ntwot 
supersymmetry this leads us to the following  bosonic action of $T_{(2,2)}$:
\beqn
S^{\rm cl}_{(2,2)}& =& \int d^2 x \left\{
 \left|\nabla_{\alpha} n^{P}\right|^2 
 +\left|\nabla_{\alpha} n^{N}\right|^2
 +\left|\tilde{\nabla}_{\alpha} \rho\right|^2
 +\left|\nabla_{\alpha} z\right|^2
 +\frac1{4e^2}F^2_{\alpha\beta} + \frac1{e^2}\,
\left|\pt_{\alpha}\sigma\right|^2
\right.
\nonumber\\[3mm]
&+&
\left|\sigma+ m_P\right|^2 \left|n^{P}\right|^2 
+ \left|\sigma+ m_{N}\right|^2\left|\rho\right|^2
+ \left|\sigma\right|^2\left|n^N\right|^2
+ \left|\sigma\right|^2\left|z\right|^2
\nonumber\\[3mm]
&+&\left.
 \frac{e^2}{2} \left(|n^{P}|^2 +|n^N|^2+|z|^2 -|\rho|^2 -2\beta\right)^2
\right\},
\nonumber\\[4mm]
&& 
P=1,...,N-1\,.
\label{Tcl}
\eeqn
The physical meaning of the $n^N$ and $z$ fields  is related to ``unwinding'' of the $(N-1)$-th string
into the $N$-th string which is, in fact,  absent, see Sec.~\ref{nastrings} and Fig.~\ref{figdipole}. 
The coefficient $b$ of this WCP model is equal to the sum of the charges of all charged fields,
\beq
b= (N-1) - 1 +1 +1 =N,
\label{b}
\eeq
where the first contribution comes from $(N-1)$ $n^P$ fields, the second one  comes
 form the size modulus $\rho$ and the last two contributions come from the ``unwinding'' fields $n^N$ and $z$.

In Sec. \ref{quantum} we will see  that
(\ref{Tcl})   is actually the action of our 2D theory at the {\em classical level}. At the quantum level the model will be 
modified by bulk quantum corrections. The superscript in $S^{\rm cl}_{(2,2)}$ reflects this.

As a check let us  choose one of  $(N-1)$ vacua of the theory (\ref{Tcl}),
\beq
n^P=\sqrt{2\beta}\;\delta^{PP_0},\qquad \sigma=-\,m_{P_0},\qquad P=1,..., N-1\,,
\label{Tclvac}
\eeq
where $P_0$ can be chosen arbitrarily from the set $\{1,..., N-1\}$.
Then in the quasiclassical limit (\ref{cllimit}) the fields $n^P$  with $P\neq P_0$ have masses $$|m_P-m_{P_0}|\,,$$
the field $\rho$ has mass  $|m_N-m_{P_0}|$, while the fields $n^N$ and $z$ are much heavier, their  mass  is $|m_{P_0}|$. Therefore  $n^N$ and $z$ can be integrated 
out which leads us to the low-energy  world-sheet theory which describes non-Abelian 
strings in the  bulk theory with the gauge group (\ref{legroup}) in the quasiclassical approximation
(\ref{cllimit}).

\begin{figure}[h]
\epsfxsize=10cm
\centerline{\epsfbox{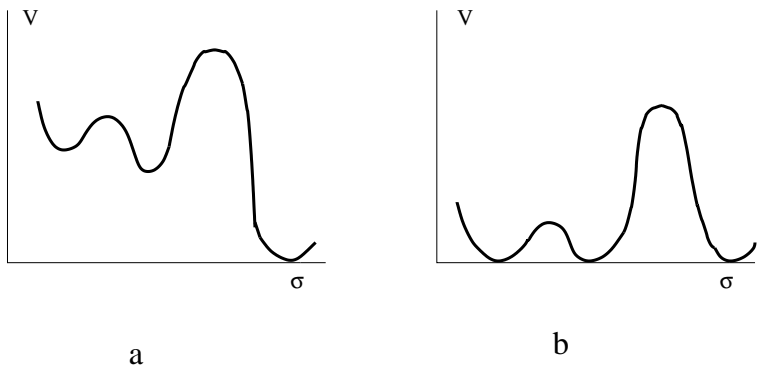}}
\caption{\small a. Schematic picture of the scalar potential in the theory (\ref{S2Dr=N-1})
in the  quasiclassical limit (\ref{cllimit}).
Now the ``last"  vacuum has zero energy reflecting the absence of the $N$-th string.
b. The same potential in the limit $\mu=0$.
}
\label{pot}
\end{figure}

Qualitative behavior of the total scalar potential in (\ref{S2Dr=N-1}) is shown in Fig.~\ref{pot}\,a.
Barriers between different strings are described by the scalar potential in (\ref{Tcl}). The heights
of these barriers are of the order of the quark mass differences squared, $|m_P-m_{P+1}|^2$. 
The height of the last barrier
 associated with the metastability of the $(N-1)$-th  string is of the order of $|m_{N-1}|^2$. As was already mentioned,
the vacuum energies at the minima are proportional to $\mu$ and are given by string tensions in Eq. (\ref{vacua}).
(They are not reflected in (\ref{Tcl}).)
The last $N$-th ``vacuum'' at  $\sigma_N = 0$ has zero energy reflecting the absence of the $N$-th string.
 Figure~\ref{pot}\,b shows the same potential in the limit of $\mu=0$. All vacua become stable and the
BPS-saturated  kinks
become well defined.

To conclude this section let us discuss a more general setup with $N_f>N$, i.e. we will add more quark 
flavors in the bulk theory. On the string world sheet
this leads to emergence of extra size moduli $\rho$. The theory $T_{(2,2)}$ is still given by the 
WCP model similar to the one in (\ref{Tcl}) with   $N-1$
orientational moduli  $n^P$ with charge +1  ($P=1,..., N-1)$) and $(N_f-N+1)$  size moduli $\rho^K$,
with charge $-1$ ($K=N,..., N_f$ ) plus two fields fields $n^N$ and $z$ with charges $+1$. The first 
coefficient of the $\beta$ function
of this world sheet theory  
\beq
b=(N-1) -(N_f-N+1) +1+1= 2N-N_f
\label{bsemilocal}
\eeq
coincides with the coefficient in the bulk theory.

\section {Quantum deformation}
\label{quantum}
\setcounter{equation}{0}

\subsection{Superpotential with quantum corrections}

The exact twisted superpotential for the classical version of the theory $T_{(2,2)}$ (see (\ref{Tcl})) is
\beqn
 &&{\cal W}^{\rm cl}= 
\frac1{4\pi}\left\{\sum_{P=1}^{N-1}\,
\left(\sigma+{m}_P\right)
\,\ln{\frac{\sigma+{m}_P}{e\Lambda}}
\right.
\nonumber\\[3mm]
&& 
\left.
+2 \sigma\,\ln{\frac{\sigma}{e\Lambda}}
-\sum_{K=N}^{N_F}\,\left(\sigma+{m}_K\right)
\,\ln{\frac{\sigma+{m}_K}{e\Lambda}} 
 \right\}\, ,
\label{clsup}
\eeqn
where the first term comes from integrating out $(N-1)$ orientational moduli $n^P$, the second term comes 
from the ``unwinding''
fields $n^N$ and $z$ and the last term comes from $N_f- N+1$ size moduli $\rho^K$. Here we generalize 
(\ref{Tcl}) adding more flavors, so that $N_f\ge N$, see the discussion at the end of the previous section.

We can rewrite (\ref{clsup}) identically in the form
\beqn
&&{\cal W}^{\rm cl}(\sigma)= 
\frac1{4\pi}\left\{
2\,{\rm Tr}\left[ (\sigma -\sqrt{2}\,\Phi^{\rm cl})\,\ln{\frac{\sigma-\sqrt{2}\,\Phi^{\rm cl}}{e\Lambda}}
\right]
\right.
\nonumber\\
&-&
\left.
\sum_{A=1}^{N_f}\,
\left(\sigma+{m}_A\right)
\,\ln{\frac{\sigma+{m}_A}{e\Lambda}}
\right\},
\label{anotherclsup}
\eeqn
where 
\beq
\Phi^{\rm cl} =-\frac1{\sqrt{2}}\,
\left(
\begin{array}{cccc}
m_1 & \ldots & 0 & 0 \\
\ldots & \ldots & \ldots  & 0 \\
0 & \ldots & m_{N-1} & 0 \\
0 & \ldots & 0  & 0
\end{array}
\right),
\label{avevr=N-1}
\eeq
see (\ref{avevr}) with $r=N-1$.

Below we will prove that, unlike the $r=N$ vacuum \cite{SYfstr}, nonperturbative effects in the bulk (in the form of the gluino condensate) affect the target space of the world-sheet model in the case $r=N-1$.
In the former case, $r=N$, the gluino condensate vanishes.
 
In the spirit of the approach put forward recently by Gaiotto, Gukov and Seiberg \cite{GGS}
we argue that the exact twisted superpotential of the world-sheet theory   $T_{(2,2)}$ 
is 
\beqn
&&{\cal W}(\sigma) = 
\frac1{4\pi}\left\{
2\,\left\langle {\rm Tr}\left[ (\sigma -\sqrt{2}\,\Phi)\,
\ln{\frac{\sigma-\sqrt{2}\,\Phi}{e\Lambda}}\right]\right\rangle
\right.
\nonumber\\
&-&
\left.
\sum_{A=1}^{N_f}\,
\left(\sigma+{m}_A\right)
\,\ln{\frac{\sigma+{m}_A}{e\Lambda}}
\right\},
\label{qusup}
\eeqn
where the braces imply that the quantum average is taken over the bulk theory.

Following \cite{GGS} we calculate the second derivative of the quantum average in (\ref{qusup})
with respect to $\sigma$ to obtain the resolvent
\beq
T(\sigma)=\left\langle {\rm Tr}\,\frac1{\sigma-\sqrt{2}\,\Phi}\right\rangle .
\label{resodef}
\eeq
The exact solution for this  object was  found  by Cachazo, Seiberg and Witten \cite{Cachazo2} precisely in our 
bulk theory.
In particular, for the bulk deformation (\ref{msuperpotbr}) in the $r$ vacuum we have 
\beqn
&& T(\sigma)_{r} =
\frac12\,\sum_{A=1}^{N_f}\,\frac{1}{\sigma+{m}_A} + \frac12\, \frac{2N-N_f}{\sqrt{\sigma^2-\frac{4S}{\mu}}}
\nonumber\\
&&-
\frac12\,\sum_{A=1}^{r}\,\frac{\sqrt{m_A^2-\frac{4S}{\mu}}}{\sqrt{\sigma^2-\frac{4S}{\mu}}(\sigma+{m}_A)}
+ \frac12\,\sum_{A=r+1}^{N_f}\,\frac{\sqrt{m_A^2-\frac{4S}{\mu}}}{\sqrt{\sigma^2-\frac{4S}{\mu}}(\sigma+{m}_A)},
\label{resolvent}
\eeqn
where $S$ is the gaugino condensate. Note that the ratio $S/\mu$ depends only on masses and $\Lambda$,
it does not depend on $\mu$, see (\ref{eN}) demonstrating that $S\propto\mu$.

In the $r=N$ vacuum the gaugino condensate is zero, $S=0$. In this case the resolvent $T(\sigma)$ in 
(\ref{resolvent}) reduces to its classical expression
\beq
T(\sigma)_{r=N}= \sum_{P=1}^{N}\,\frac{1}{\sigma+{m}_P}.
\label{resolventrN}
\eeq
This gives the superpotential (\ref{CPsup}) for the theory on non-Abelian string in the $r=N$ vacuum. 
Quantum deformation is absent in this case. Since the superpotential  (\ref{CPsup}) contains only
one-loop logarithmic terms we can say that bulk instantons do not penetrate on the world sheet in 
the $r=N$ vacuum.

Consider now the $r=N-1$ vacuum.
Integrating over $\sigma$ in (\ref{resolvent}) and substituting the result
 into (\ref{qusup}) we  get
\beqn
&&\partial_{\sigma}{\cal W}(\sigma)=
\frac1{4\pi}\left\{
\sum_{A=1}^{N-1}\,\ln{\frac{(\sigma+{m}_A)}{\Lambda}}-\sum_{A=N}^{N_f}\,\ln{\frac{(\sigma+{m}_A)}{\Lambda}}
\right.
\nonumber\\
&+&
\left.
(2N-N_f)\,\ln{\frac{t}{\Lambda}}
-\sum_{A=1}^{N-1}\,\ln{\frac{t_A}{\Lambda}}
+\sum_{A=N}^{N_f}\,\ln{\frac{t_A}{\Lambda}}
\right\},
\label{dW}
\eeqn
where we define
\beq
t=\frac12\left(\sigma +\sqrt{\sigma^2-\frac{4S}{\mu}}\right)
\label{t}
\eeq
and 
\beq
t_A=\frac12\left(\sqrt{\sigma^2-\frac{4S}{\mu}}+
 \frac{\sigma +\frac{4S}{\mu m_A}}{\sqrt{1-\frac{4S}{\mu m_A^2}}}\right).
\label{tA}
\eeq

\vspace{2mm}

Equation~(\ref{dW}) is our final result for the exact twisted superpotential of the theory $T_{(2,2)}$ 
on the non-Abelian string in the $r=N-1$
vacuum. It takes into account the quantum deformation produced by the bulk instantons. The latter 
generate gaugino condensate which results in the emergence of the square root cut in the
$\sigma$ plane, see (\ref{t}) and (\ref{tA}). The emergence of this  cut is a response of the 2D world-sheet theory 
 to the cut in the SW curve present in the bulk theory in the $r=N-1$ vacuum, see (\ref{rcurve}).

If we neglect the gaugino condensate in the quasiclassical limit of large masses we get
$t\approx t_A \approx \sigma$ returning us to the ``classical'' superpotential (\ref{clsup}).

\subsection{Chiral ring equation}

The equation determining the vacuum values of $\sigma$ is obtained by requiring $\partial_{\sigma}{\cal W}(\sigma)=0$.
Exponentiating (\ref{dW}) we obtain
\beq
t^{(2N-N_f)}\,\prod_{P=1}^{N-1}\frac{(\sigma+{m}_P)}{t_P}
=\Lambda^{(2N-N_f)}\,\prod_{K=N}^{N_f}\frac{(\sigma+{m}_K)}{t_K}\,.
\label{qusigmaeq}
\eeq

Let us first approximately solve this equation  in the quasiclassical limit of large (generic) quark masses,
\beq
m_A\gg\Lambda, \qquad \Delta m_{AB} \sim  m_A.
\label{cllimitg}
\eeq
To the leading order $N-1$ roots are 
$$\sigma_P\approx -m_P\,\qquad P=1,..., \,N-1\,.$$
 Consider the first 
correction to a particular root $\sigma_{P_0}$ where $P_0$ is arbitrarily chosen  from the set $\{1,...,  N-1\}$.  From Eq.~(\ref{qusigmaeq}) we 
find
\beq
\sigma_{P_0}\approx -m_{P_0} + \frac{\Lambda^{(2N-N_f)}}{m_{P_0}^2}\;
\frac{\prod_{K=N}^{N_f}({m}_K-m_{P_0})}{\prod_{P\neq P_{0}}({m}_P-m_{P_0})} + \cdots,
\label{sigmaP}
\eeq
where the product in the denominator runs over $P=1,...,\, N-1$ and 
we neglect the gaugino condensate as compared to the quark masses.

In Appendix A we present the calculation of the double roots of the SW curve in the $r=N-1$ vacuum with
the same accuracy. Comparing (\ref{sigmaP}) and  (\ref{eP}) we see that
\beq
\sigma_P=\sqrt{2}\,e_P, \qquad P=1,...,\, N-1\,.
\label{equalrootsP}
\eeq

Now let us calculate ``small'' roots in (\ref{qusigmaeq}). We will see that they will be of the
order of $\sqrt{S/\mu} \sim e_N$, see (\ref{eN}). Therefore we still can neglect $S/\mu$ as compared to
$\sigma$ or $m_A$. This gives $t_A\approx t$, and then Eq.~(\ref{qusigmaeq}) reduces to
\beq
t^2\,\prod_{P=1}^{N-1}(\sigma+{m}_P)
=\Lambda^{(2N-N_f)}\,\prod_{K=N}^{N_f}(\sigma+{m}_K)\,.
\label{qusigmaeqapp}
\eeq
Neglecting small $\sigma$ as compared to the quark masses we get
\beq
t^2 \approx \Lambda^{(2N-N_f)}\,\frac{\prod_{K=N}^{N_f}{m}_K}{\prod_{P=1}^{N-1}{m}_P}, \qquad
t\approx \pm\sqrt{\Lambda^{(2N-N_f)}\,\frac{\prod_{K=N}^{N_f}{m}_K}{\prod_{P=1}^{N-1}{m}_P}}\,.
\label{teq}
\eeq
The values of the unpaired roots of the SW curve $e_N^{\pm}$ are calculated in Appendix A in the leading order. 
  Equation~(\ref{eNapp}) shows that the combination under the square root sign in (\ref{teq}) is exactly
$e_N^2/2$. Then the above equation gives
\beq
2t=\left(\sigma +\sqrt{\sigma^2-2e_N^2}\right)=\sqrt{2}\, e_N^{\pm}\,,
\eeq
where we use (\ref{t}) and (\ref{eN}).

From this equation we find  that two ``small'' roots $\sigma^{\pm}$ are
\beq
 \sigma_N^{\pm}\approx \pm 2\sqrt{\Lambda^{2N-N_f}
\, \frac{\prod_{K=N}^{N_f}\, m_K}{\prod_{P=1}^{N-1} \, m_P}}\,.
\label{sigmaN}
\eeq
They are given by unpaired roots of the SW curve,
\beq
\sigma_N^{\pm}=\sqrt{2}\,e_N^{\pm}\,.
\label{equalrootsN}
\eeq

Thus, we see that VEVs of $\sigma$ are given by the roots of the SW curve in 
the $r=N-1$ vacuum in much the same way
 as   is the case in the $r=N$ vacuum. We proved this statement in the quasiclassical approximation above. 
However, it is very likely that this relation is exact. We assume this conjecture to be true in what follows. 

The emergence of two roots $\sigma_N^{\pm}$ is a reflection of the cut in the $\sigma$ plane.
Classically the cut is invisible.

\section {2D-4D correspondence in the \boldmath{$r=N-1$} vacuum}
\label{2D/4D}
\setcounter{equation}{0}

Since confined monopoles are represented by kinks in the world-sheet theory  their masses 
should coincide. In this section we explicitly confirm this expectation by
verifying the equality in  (\ref{equalmasses}).

\subsection{Kink masses versus monopole masses}

If we neglect non-BPS $\mu$-corrections in (\ref{kinkmassgen}) the kink masses    
 are 
\beq
M^{\rm kink}_{PP'} =  2\left|{\cal W}(\sigma_{P'})-{\cal W}(\sigma_{P})\right|,
\qquad P,P'=1,...,N,
\label{kinkmassg}
\eeq
where ${\cal W}(\sigma)$ for $T_{(2,2)}$ is determined by (\ref{dW}). Starting from  $\partial_\sigma{\cal W}(\sigma)$
from (\ref{dW}) and integrating by parts we can present the kinks masses in the form
\beqn
&&  M^{\rm kink}_{PP'} = \left|\frac{1}{\pi}\int_{\sigma_P}^{\sigma_{P'}}\, d\sigma \left\{
 \frac{2N-N_f}{2}\, \frac{\sigma}{\sqrt{\sigma^2-\frac{4S}{\mu}}}
\right.\right.
\nonumber\\
&-&
\left.\left.
\frac12\,\sum_{A=1}^{N-1}\,\frac{\sigma\sqrt{m_A^2-\frac{4S}{\mu}}}
{\sqrt{\sigma^2-\frac{4S}{\mu}}\,(\sigma+{m}_A)}
+ \frac12\,\sum_{A=N}^{N_f}\,\frac{\sigma\sqrt{m_A^2-\frac{4S}{\mu}}}
{\sqrt{\sigma^2-\frac{4S}{\mu}}\,(\sigma+{m}_A)}
\right\}\right|,
\nonumber\\
\label{Mkinkint}
\eeqn
where we drop the total derivative term.  It is zero for the vacuum values of $\sigma$ due to
 Eq.~(\ref{qusigmaeq}).

In Appendix B we present  calculation of the monopole mass in the $r=N-1$ vacuum for the simplest example: U(2) theory
with two flavors, $N_f=2$. Taking $N=N_f=2$ in (\ref{Mkinkint}) we see that the kink mass coincides 
 with the mass (\ref{Zmon}) of the monopole.  The important input here is the coincidence
of the roots of the SW curve with the vacuum values of $\sigma$, see (\ref{equalrootsP}) and (\ref{equalrootsN}).

\subsection{Kink mass in U(2)}

For illustration we calculate the kink/monopole mass  in the $r=N-1$ vacuum in the simplest U(2) theory
with $N_f=2$ in the semiclassical approximation
\beq
m_1\sim m_2 \gg \Lambda\,.
\label{cllimitU2}
\eeq
In this case Eq.~(\ref{dW}) reads
\beq
\partial_{\sigma}{\cal W}(\sigma)\approx
\frac1{4\pi}\left\{
\ln{\frac{(\sigma+{m}_1)}{\Lambda}}-\ln{\frac{(\sigma+{m}_2)}{\Lambda}}
+2\,\ln{\frac{t}{\Lambda}}\right\},
\label{dWU2}
\eeq
where we used $t_A\approx t$ in the semiclassical approximation (\ref{cllimitU2}), see (\ref{t}) and (\ref{tA}). 
Integrating over $\sigma$ 
we get
\beq
M^{\rm kink}\approx \left|\frac1{2\pi}\left\{
m_1\ln{(\sigma +m_1)} -m_2\ln{(\sigma +m_2)}-2\sqrt{\sigma^2-4\Lambda^2\,\frac{m_2}{m_1}}
\right\}^{\sigma_2^{\pm}}_{\sigma_1}\right|
\rule{0mm}{10mm}
,
\\
\label{ZkinkU2m}
\eeq
Here we used (\ref{eNapp}) to calculate $4S/\mu$, see (\ref{eN}).
In (\ref{ZkinkU2m}) the kink central charge is given by the difference of the expression in the braces
 calculated at $\sigma=\sigma_2^{\pm}$ and  $\sigma=\sigma_1$, respectively,  We drop the term proportional to $\sigma$ 
since it vanishes due to (\ref{qusigmaeqapp}). As usual, different branches of the logarithmic functions 
will give dyonic kinks.

Equations  (\ref{sigmaP}) and (\ref{sigmaN})  
imply
\beq
\sigma_1\approx -m_1 +(m_2-m_1)\,\frac{\Lambda^2}{m_1^2},
\qquad \sigma_2^{\pm}\approx \pm 2\Lambda \,\sqrt{\frac{m_2}{m_1}}\,.
\label{sigmavevU2}
\eeq
Substituting this in (\ref{ZkinkU2m}) we finally get
\beq
M^{\rm kink} =\left|\frac1{\pi}\left\{
m_1\,\ln{\frac{m_1}{\Lambda}} +\frac{m_1}{2}\,\ln{\frac{m_1}{m_2-m_1}}
-\frac{m_2}{2}\,\ln{\frac{m_2}{m_2-m_1}} +m_1 +\cdots\right\}\right|,
\label{ZkinkU2}
\eeq
where the ellipses denote terms proportional to $\Lambda$. Note that the result for the kink mass
 does not depend on the particular choice of the upper limit\,\footnote{This statement can be proven to be exact.} 
 either $\sigma=\sigma_2^{+}$ or $\sigma=\sigma_2^{-}$.

The coefficient in front of the logarithm of $\Lambda$ here is $b/2\pi$; it reflects the correct coefficient of the
$\beta$ function, $b=2$.

It is instructive to consider different limits in (\ref{ZkinkU2}). First take the limit $m_2\to\infty$
decoupling the second flavor. In this case the coefficient of the $\beta$ function $b$ becomes
$b_1=3$ while the effective scale of the theory is $\Lambda_1^3 =m_2\Lambda^2$. Eq.~(\ref{ZkinkU2})
gives in this limit
\beq
M^{\rm kink}_{N_f=1} =\left|\frac1{\pi}\left\{
\frac32\,m_1\,\ln{\frac{m_1}{\Lambda_1}}  +\frac12\,m_1 +\cdots\right\}\right|.
\label{ZkinkU21f}
\eeq
We see that the logarithmic term here has the correct coefficient $b_1/2\pi=3/2\pi$ 
in front of $\ln{\Lambda_1}$.

Another interesting limit is that of the equal masses $\Delta m =m_2-m_1\to 0$. In this limit (\ref{ZkinkU2})
reduces to
\beq
M^{\rm kink} =\left|\frac1{\pi}\left\{
m\,\ln{\frac{m}{\Lambda}} -\frac{\Delta m}{2}\,\ln{\frac{m}{\Delta m}}
 +m +\cdots\right\}\right|,
\label{ZkinkU2appeqmass}
\eeq
where $m_2\approx m_1=m$. 
We see that the kink mass stays finite in the limit $\Delta m\to 0$ as expected.

In fact, it is possible to obtain an exact formula for the kink mass in the limit $m_1=m_2$.
Using (\ref{dW}) it is easy to show that in the limit $\Delta m\to 0$ the kink mass is still 
given by Eq.~(\ref{ZkinkU2m}).
To calculate singular logarithmic terms in this expression Eq.~(\ref{sigmavevU2}) should be modified.
Using (\ref{qusigmaeq}) we get
\beq
\sigma_1 =  -m +\Delta m\,\frac{2\Lambda^2}{(m+\sqrt{m^2-4\Lambda^2})\sqrt{m^2-4\Lambda^2}} +O(\Delta m^2),
\qquad \sigma_2^{\pm} = \pm 2\Lambda \,,
\label{sigmavevU2exact}
\eeq
where now we do not assume that $m\gg \Lambda$.

Substituting this in (\ref{ZkinkU2m}) we get the exact result
\beq
M^{\rm kink} =\left|\frac1{2\pi}\left\{
m\,\ln{\frac{m+\sqrt{m^2-4\Lambda^2}}{m-\sqrt{m^2-4\Lambda^2}}} 
 +2\sqrt{m^2-4\Lambda^2} \right\}\right|.
\label{ZkinkU2eqmass}
\eeq

For comparison we quote the kink mass on the non-Abelian string  in the U(2) gauge theory with $N_f=2$
in the $r=2$ vacuum \cite{Dorey},
\beqn
&&
M^{\rm kink}_{r=N}=\left|\frac1{2\pi}\left\{
\Delta m\,\ln{\frac{\Delta m+\sqrt{\Delta m^2+4\Lambda^2}}{\Delta m-\sqrt{\Delta m^2+4\Lambda^2}}} 
 -2\sqrt{\Delta m^2+4\Lambda^2} \right\}\right| 
\nonumber\\
&=&\left|\frac1{\pi}\left\{
\Delta m\,\ln{\frac{\Delta m}{\Lambda}}  -\Delta m\right\} +\frac{i}{2}\Delta m +\cdots\right|.
\label{ZkinkrN}
\eeqn
We see that the kink masses on the strings in the two vacua are different. Both kink masses coincide with 
the bulk monopole masses in the corresponding vacua. 

\section {Deformation potential for the world sheet theory}
\label{deformation}
\setcounter{equation}{0}

Now let us discuss the $\mu$-dependent deformation potential (the second term in (\ref{S2Dr=N-1})) in the theory on the
 non-Abelian string in  the $r=N-1$ vacuum. 
An obvious modification of the $r=N$  world-sheet potential (\ref{Vdefr=N}) is 
\beq
 V_{\rm def} (\sigma)=4\pi\,\left|\mu\,\sqrt{\sigma^2-\frac{4S}{\mu}}\right|.
\label{def}
\eeq
This potential correctly reproduces string tensions in the vacua determined by Eq.~(\ref{qusigmaeq}),
see (\ref{vacua}). Quasiclassically these vacua are given by (\ref{sigmaP}) and (\ref{sigmaN}).
 At two points $\sigma_N^{\pm}$ vacuum energy is zero. This ``vacuum'' corresponds to the non existing
$N$-th string. The split of this ``vacuum,'' which classically corresponds to  $\sigma_N \approx 0$ is
due to the cut which opens up on the sigma plane. Classically this cut is invisible.

As was already mentioned, the deformation potential (\ref{def}) breaks \ntwot supersymmetry down to
\ntwoo which is further spontaneously broken  by choosing a vacuum with nonvanishing energy.

\section {Conclusions}
\label{concl}
\setcounter{equation}{0}

In this paper we continue explorations of the \ntwoo theories emerging on 
non-Abelian strings supported by  \ntwo supersymmetric QCD
with the gauge group U$(N)$ and $N_f$ flavors of quarks ($N_f\geq N$). 
 \ntwo supersymmetry
is broken down to \none by a small deformation: a small mass
term for the adjoint matter. 
The beginning of this program was reported in Ref. \cite{SYfstr} in which non-Abelain strings were considered
in the $r=N$ vacuum. (Remember, $r$ is the number of condensed (s)quarks in the large quark mass limit.)

In \cite{SYfstr} we obtained a weighted CP model on the string world sheet. 
Nonperturbative corrections which determine the BPS spectrum could be
derived ``inside" this two-dimensional model {\em per se}.

We discover that 
the situation drastically changes in passing from $r=N$ to $r=N-1$.
In the $r=N-1$ case which is our main focus, 
the low-energy two-dimensional theory on the string world-sheet
receives nonperturbative corrections from the bulk, through the bulk gaugino condensate. 
Classically the world-sheet theory is still a weighted CP model. However, a nonvanishing
gluino condensate in the bulk affects the target space on the world sheet, generating additional nonperturbative effects that lie ``outside" the original classical model. We calculated these additional nonperturbative  effects deforming the weighted CP model on the world sheet 
 by virtue of the method of resolvents suggested by Gaiotto, Gukov 
and Seiberg for surface defects \cite{GGS}.

The 2D-4D correspondence (the coincidence of spectra of two-dimensional kinks and four-dimensional monopoles)
remains valid in the BPS sector. 
 
The target space deformation in the world-sheet model after
penetration of the bulk corrections remains unidentified. We managed to derive the exact twisted two-dimensional  superpotential bypassing this stage. Such an identification is a task for the future.
Another  obvious direction for future work is   generalization of the construction presented here to  strings
in the generic $r$ vacua of $\mu$-deformed \ntwo QCD.

\section*{Acknowledgments}
This work  is supported in part by DOE grant DE-FG02-94ER40823. 
The work of A.Y. was  supported 
by  FTPI, University of Minnesota, 
by RFBR Grant No. 13-02-00042a 
and by Russian State Grant for 
Scientific Schools RSGSS-657512010.2.

\section*{Appendix A:  \\
Roots of the Seiberg-Witten curve in the\\
 \boldmath{$r=N-1$} vacuum}
\label{appA}
\addcontentsline{toc}{section}{Appendices}

 \renewcommand{\theequation}{A.\arabic{equation}}
\setcounter{equation}{0}
 
 \renewcommand{\thesubsection}{A.\arabic{subsection}}
\setcounter{subsection}{0}

To identify the $r=N-1$ vacuum in terms of the SW curve 
\beq
y^2= \prod_{P=1}^{N} (x-\phi_P)^2 -
4\left(\frac{\Lambda}{\sqrt{2}}\right)^{2N-N_f}
\, \,\,\prod_{A=1}^{N_f} \left(x+\frac{m_A}{\sqrt{2}}\right)
\label{curveA}
\eeq
we have to find $\phi$ such that the curve factorizes as in (\ref{rcurve}) and at large quark masses
$(N-1)$ values of $\phi_P$ are determined by masses, see (\ref{classphi}).

Let us calculate $\phi_P$'s and the roots $e_P$'s in the quasiclassical approximation (\ref{cllimitg}).
First consider ``large'' 
$$
\phi_{P}\approx -m_P/\sqrt{2}\,,\qquad P=1,...,(N-1)\,.$$
 Let us calculate the 
correction to a particular  $\phi_{P_0}$, $P_0=1,...,(N-1)$. The SW curve reduces to
\beq
y^2\sim (x-\phi_{P_0})^2 - 
4\left(\frac{\Lambda}{\sqrt{2}}\right)^{2N-N_f}
\, \frac{\prod_{K=N}^{N_f} \frac{m_K-m_{P_0}}{\sqrt{2}}}{\prod_{P\neq P_{0}} \frac{m_P-m_{P_0}}{\sqrt{2}}}
\,\frac{2}{m_{P_0}^2}\,  \left(x+\frac{m_{P_0}}{\sqrt{2}}\right),
\label{Pcurve}
\eeq
where $P=1,...,N-1$.
Now we look for $\phi_{P_0}$ which ensures that the corresponding root $e_{P_0}$ is a double root.
We then get
\beqn
 \sqrt{2}\,\phi_{P_0}&=& -m_{P_0} -  \Lambda^{2N-N_f}
\, \frac{\prod_{K=N}^{N_f} (m_K-m_{P_0})}{\prod_{P\neq P_{0}} (m_P-m_{P_0})}
\,\frac{1}{m_{P_0}^2} + \cdots\,,
\nonumber\\[3mm]
\sqrt{2}\,e_{P_0}&=& -m_{P_0} +  \Lambda^{2N-N_f}
\, \frac{\prod_{K=N}^{N_f} (m_K-m_{P_0})}{\prod_{P\neq P_{0}} (m_P-m_{P_0})}
\,\frac{1}{m_{P_0}^2} +\cdots\,.
\label{eP}
\eeqn
The value of the double root $\sqrt{2}\,e_{P_0}$ here coincides with the VEV $\sigma_{P_0}$ 
(see (\ref{sigmaP}))
calculated from the world-sheet theory.

Now consider ``small'' unpaired roots $e_N^{\pm}$. Assuming that $x$ is small  compared to the quark masses
we obtain for the SW curve
\beq
y^2\sim (x-\phi_{N})^2 - 
4\left(\frac{\Lambda}{\sqrt{2}}\right)^{2N-N_f}
\, \frac{\prod_{K=N}^{N_f} \frac{m_K}{\sqrt{2}}}{\prod_{P=1}^{N-1} \frac{m_P}{\sqrt{2}}}
\,.
\label{Ncurve}
\eeq
The condition (\ref{DijVafa}) gives $\phi_N\approx 0$ and 
\beq
\sqrt{2}\, e_N^{\pm}\approx \pm 2\sqrt{\Lambda^{2N-N_f}
\, \frac{\prod_{K=N}^{N_f}\, m_K}{\prod_{P=1}^{N-1} \, m_P}}.
\label{eNapp}
\eeq
This result shows that the combination under the square root in (\ref{teq})  is precisely
$e_N^2/2$. Then (\ref{equalrootsN}) follows.

\section*{Appendix B:  \\
Monopole mass in the \boldmath{$r=N-1$} vacuum}
\label{appB}
\addcontentsline{toc}{section}{Appendices}

 \renewcommand{\theequation}{B.\arabic{equation}}
\setcounter{equation}{0}
 
 \renewcommand{\thesubsection}{B.\arabic{subsection}}
\setcounter{subsection}{0}

The monopole mass is given by
\beq
M^{\rm monopole}_{PP'}=\left|\frac{\sqrt{2}}{2\pi i}\,\oint_{\beta_{PP'}} d\lambda_{SW}\right|,
\qquad P,P'=1,....N,
\label{Zmongen}
\eeq
see (\ref{monmassgen}). Here the SW differential is defined as \cite{HaOz,GMMM}
\beq
d\lambda_{SW}=\frac{xdP}{y} - \frac{xP}{2y}\,\frac{dQ}{Q}+\frac{x}{2}\,\frac{dQ}{Q}
\label{dlambda}
\eeq
where $P$ and $Q$ are polynomials which enter the SW curve (\ref{curve}),
\beq
P(x)= \prod_{P=1}^{N} (x-\phi_P), \qquad Q(x)= \prod_{A=1}^{N_f} \left(x+\frac{m_A}{\sqrt{2}}\right).
\eeq
The residues at $x=-m_A /\sqrt{2}$ in (\ref{dlambda}) are fixed to be integers of $m_A$.
Moreover, the SW differential (\ref{dlambda}) does not have poles at $x=e_P$.

Consider the simplest case of the U(2) theory with $N_f=2$. The SW curve in the $r=N-1=1$ vacuum factorizes as 
in (\ref{rcurve}). The monopole mass takes the form
\beqn
&&  M^{\rm monopole} = \left|\frac{1}{\pi}\int_{\sqrt{2}e_1}^{\sqrt{2}e_2^{\pm}}\, zdz \left\{
  \frac{1}{\sqrt{z^2-2e_2}}
\right.\right.
\nonumber\\[4mm]
&-&
\left.\left.
\frac12\,\frac{\sqrt{m_1^2-2e_2^2}}
{\sqrt{z^2-2e_2^2}\,(z+{m}_1)}
+ \frac12\,\frac{\sqrt{m_2^2-2e_2^2}}
{\sqrt{z^2-2e_2^2}\,(z+{m}_2)}
\right\}\right|,
\label{Zmon}
\eeqn
where the integration variable $z=\sqrt{2}\,x$. We see that the integral representations for the four-dimensional
 monopole and two-dimensional kink masses
are the same, see (\ref{Mkinkint}).

\end{document}